\documentclass[runningheads]{llncs}
\input epsf

\begin{document}

\pagestyle{headings}

\mainmatter
\title{Correlation between mutation pressure, \\
selection pressure\\
 and occurrence of amino acids}
 
\titlerunning{Correlation between mutation pressure, selection pressure}

\author{Aleksandra Nowicka\inst{1}
\and Pawe{\l} Mackiewicz\inst{1} 
\and Ma{\l}gorzata Dudkiewicz\inst{1}
\and Dorota Mackiewicz\inst{1}
\and Maria Kowalczuk\inst{1} 
\and Stanis{\l}aw Cebrat\inst{1} 
\and Miros{\l}aw R. Dudek\inst{2}\thanks{corresponding author}
}  

\authorrunning{Aleksandra Nowicka et al.} 

\institute{Department of Genetics, Institute of Microbiology,
University of Wroclaw,\\
ul. Przybyszewskiego 63/77, PL-54148 Wroclaw, Poland\\
\email{\{nowicka, pamac, malgosia, dorota, kowal,
cebrat\}@microb.uni.wroc.pl}\\
\texttt{http://smORFland.microb.uni.wroc.pl}
\and
Institute of Physics, University of Zielona G{\'o}ra,
ul. A. Szafrana 4a, \\
PL-65516 Zielona G{\'o}ra, Poland\\
\email{mdudek@proton.if.uz.zgora.pl}\\
}

\maketitle

\begin{abstract}
We have found that the effective survival time of  amino acids
 in organisms follows a power law with respect to frequency of 
 their occurrence in genes.
 We have used mutation data matrix PAM$1$ PET$91$ 
 to calculate selection pressure
 on each kind of amino acid. The results have been compared to 
 MPM$1$ matrix (Mutation Probability Matrix) 
 representing the pure mutational pressure in the {\it Borrelia burgdorferi} genome.
 The results are universal in the sense that the survival time
 of amino acids calculated from 
 the higher order PAM$k$ matrices ($k>1$)
  follows the same power law as  in the case of PAM$1$ matrices. 
\end{abstract}

\section{Introduction}
Determining the evolutionary distances between 
two protein sequences requires
the knowledge of the substitution rates of amino-acids. It is generally 
accepted that the more substitutions are necessary to change one sequence 
into another, the more unrelated they are and the larger their distance
to the common ancestor. The most widely used method for the calculation of distances between sequences is based on the mutation data matrix, $M_{ij}$,
 published by Dayhoff et
al. \cite{l_Dayhoff}, where $i,j$ represent amino acids, and
an element $M_{ij}$ of the matrix gives the probability that the amino acid in
column $j$ will be replaced by the amino acid in row $i$ after a given evolutionary
time interval. The interval corresponding to $1$ percent of substitutions between
two compared sequences is called one PAM (Percent of Accepted Mutations),
 and the corresponding matrix is denoted as PAM$1$ matrix. 
There is assumed a Markov model of sequence evolution and a simple power $M^k$
of the PAM$1$ matrix (multiplied by itself $k$ times) 
denotes a matrix, PAM$k$, 
that gives the amino acid substitution probability after $k$ PAMs. 
Today, a much more accurate PAM matrix is available, generated 
from 16130 protein sequences, published by 
Jones et al. \cite{l_Jones}. The large number of compared genes 
guarantees that the matrix has negligible statistical errors and it can be 
considered to be the reference matrix during the calculations of the phylogenetic
distances.
The matrix is also known as PET$91$ matrix. 

Recently, by comparing intergenic sequences being remnants of coding sequences with homologous sequences of genes, 
we have    
constructed an empirical table 
of the nucleotide substitution rates in the case of the leading
DNA strand of the {\it B. burgdorferi} genome \cite{l_linear1},\cite{l_linear2},\cite{l_linear3}.
We have found   
that substitution rates, which determine the evolutionary
turnover time of a given kind of nucleotide in third codon positions of coding sequences, 
are highly correlated with the 
frequency of the occurrence of that nucleotide 
in the sequences. There is a compositional
bias produced by replication process, introducing 
long-range correlation 
among nucleotides in the third positions in codons, which is very
 similar to the bias seen
in the intergenic sequences \cite{l_cebrat1}. 
   
We have used the empirical table of nucleotide substitution rates
to simulate mutational pressure 
on the genes lying on the leading DNA strand of the {\it B.burgdorferi} genome
and we have constructed  
MPM$1$  matrix (Mutation Probability Matrix) 
for amino acid substitutions in the evolving genes. 
Thus the resulting table represents the percent of amino acid substitutions
introduced by mutational pressure and not by selection.  
Next,
we compared the survival times
of the amino acids in the case without any selection 
with the effective survival times 
of the amino acids, counted with the help of 
the PAM1 PET91 matrix.      

\section{Mutation Table for Nucleotides}

DNA sequence of the {\it B.burgdorferi} genome was downloaded from the website \\
{\it www.ncbi.nlm.nih.gov}. The empirical mutation table for nucleotides in 
third positions in codons, which we used in the paper,  
is the following \cite{l_linear1},\cite{l_linear2}:

\begin{equation}
  M = \left(
 \begin{array}{llll}
  1-uW_{A} & u~W_{AT} & u~W_{AG} & u~W_{AC} \\
  u~W_{TA} & 1-uW_{T} & u~W_{TG} & u~W_{TC}\\
  u~W_{GA} & u~W_{GT} & 1-uW_{G} & u~W_{GC}\\
  u~W_{CA} & u~W_{CT} & u~W_{CG} & 1-uW_{C}
 \end{array}
\right)
\label{macierz1}
\end{equation}
\noindent
where \footnote[1]{The transpose matrix convention 
has been chosen in \cite{l_linear1}.}
\begin{equation}
\begin{array}{l}
W_{GA}=0.0667~~
W_{GT}=0.0347~~
W_{GC}=0.0470~~
W_{AG}=0.1637\\
W_{AT}=0.0655~~
W_{AC}=0.0702~~
W_{TG}=0.1157~~
W_{TA}=0.1027\\
W_{TC}=0.2613~~
W_{CG}=0.0147~~
W_{CA}=0.0228~~
W_{CT}=0.0350\\
\end{array}
\label{r_wspol}
\end{equation}

\noindent
and the elements of the matrix give the probability that nucleotide in column
$j$ will mutate to the nucleotide in row $i$ during one replication cycle. The 
symbols $W_{ij}$ represent relative substitution probability of nucleotide $j$ by 
nucleotide $i$, and $u$ 
represents mutation rate. The symbols $W_j$ in the diagonal represent relative
substitution probability of nucleotide $j$:

\begin{equation}
W_{j}=\sum_{i \neq j} W_{ij},
\end{equation}

\noindent
and $W_A+W_T+W_G+W_C=1$.

The expression for the mean survival time of the nucleotide $j$   
depends on  
$W_{j}$
 as follows (derivation can be found in \cite{l_linear1}) 

\begin{equation}
  \tau_{j}=-\frac{1}{ln(1-u~W_{j})} \approx \frac{1}{u~W_{j}}.
\label{r_approx}
\end{equation}

\noindent
The above approximated formula is true for small values of the mutation rate $u$.

In papers \cite{l_linear1},\cite{l_linear2},\cite{l_linear3}, we concluded that 
in a natural genome the frequency of occurrence $f_j$ of the nucleotides, 
in the third position in codons, 
is linearly related to the respective
mean survival time $\tau_{j}$,

\begin{equation}
f_{j} = m_0~ \tau_{j} + c_0 ,
\label{r_lin}
\end{equation}

\noindent
with the same coefficients, $m_0$ and $c_0$, for each nucleotide.
The Kimura's neutral theory \cite{l_neutral} of evolution assumes the
constancy of the evolution rate, where the mutations are random events,
much the same as the random decay events of the radioactive decay.
However, the linear law in (\ref{r_lin}) is not contrary 
to the Kimura's theory.
Still, the mutations represent random decay events but they are 
correlated with the DNA composition.

\section{Mutational Pressure MPM$1$ Matrix Construction for Amino Acids}

We used computer random number generator to generate  
random deviates with a distribution defined by the elements of the mutation 
matrix in (\ref{macierz1}). With the help of the random deviates  
 we were mutating nucleotide sequences of 564 genes from 
the leading DNA strand of the {\it B. burdorferi} genome. The applied value
of the mutation pressure was $u=0.01$.

For each gene, considered to be an ancestral one in this simulated
evolution, we prepared $10^5$ pairs 
of homologous sequences, which diverged from this gene. The gene evolution
was stopped when the number of substitutions between 
the homologous protein sequences reached 1\%. 
All the sequences were translated into amino acids
and we constructed a mutation probability matrix MPM$1$ 
according to the procedure of Dayhoff et al.\cite{l_Dayhoff} and Jones et al.
\cite{l_Jones}. The resulting mutation table, with substitution 
probabilities $M_{ij}$, the 
amino acid mutability $m_j$, 
and the fraction $f_j$ of amino-acid in the compared sequences 
have been presented, respectively, in Table \ref{table2} 
and Table \ref{mutability}.

The elements $M_{ij}$ of the MPM$1$ matrix  in Table \ref{table2} have been  
scaled with the parameter $\lambda$, which 
related them to the evolutionary distance of one
percent of substitutions and it is equal to 0.00009731 in our simulations.
We introduced the parameter  
 $\lambda$ following the equation (3)  in the paper by Jones et
al.\cite{l_Jones}). Therefore,   

\setlength{\tabcolsep}{4pt}
\begin{table}
\begin{center}
\caption{Mutation Probability Matrix for an evolutionary 
distance of 1 PAM (splitted into two parts).
Values of the matrix elements are scaled by a factor of $10^5$ and rounded to 
an integer. The
symbols in the first row and the first column represent amino-acids and numbers following colons -number of codons representing a given amino-acid in the universal genetic code.}
\label{table2}
\begin{tabular}{|l||l|l|l|l|l|l|l|l|l|l|}
\hline
 &A:4&R:6&N:2&D:2&C:2&Q:2&E:2&G:4&H:2&I:3\\
\hline
\hline
 A:4&99027&0&0&79&1&0&57&78&0&1\\
R:6&0&98773&1&0&251&139&1&218&175&51\\
N:2&0&2&98926&336&4&1&1&1&308&162\\
D:2&101&0&338&98936&4&0&297&215&184&0\\
C:2&0&50&0&0&97442&0&0&87&1&0\\
Q:2&0&76&0&0&0&99244&41&0&500&0\\
E:2&83&2&1&340&0&151&99125&247&1&0\\
G:4&92&357&1&199&721&0&200&99103&0&0\\
H:2&0&56&56&33&2&295&0&0&98315&0\\
I:3&1&153&277&1&3&0&0&1&1&99026\\
L:6&1&90&0&0&4&198&0&0&459&150\\
K:2&0&481&452&1&0&279&354&1&1&122\\
M:1&0&69&0&0&0&0&0&0&0&137\\
F:2&1&0&1&0&697&0&0&0&2&165\\
P:4&55&52&0&0&1&82&0&0&180&0\\
S:6&218&324&215&1&1070&0&0&161&2&79\\
T:4&297&46&71&0&1&0&0&0&0&163\\
W:1&0&63&0&0&190&0&0&41&0&0\\
Y:2&0&0&235&198&868&1&1&0&591&0\\
V:4&435&1&1&213&2&0&169&235&1&225\\
\hline
\end{tabular}
~\\
~\\
~\\
\begin{tabular}{|l||l|l|l|l|l|l|l|l|l|l|}
\hline
 &L:6&K:2&M:1&F:2&P:4&S:6&T:4&W:1&Y:2&V:4\\
\hline
\hline
 A:4&0&0&0&0&107&129&333&0&0&279\\
R:6&29&177&116&0&73&140&38&653&0&0\\
N:2&0&292&1&1&0&163&102&0&330&1\\
D:2&0&1&0&0&0&1&0&0&279&176\\
C:2&0&0&0&62&0&91&0&389&137&0\\
Q:2&34&56&0&0&63&0&0&1&1&0\\
E:2&0&263&1&0&0&0&0&3&1&159\\
G:4&0&1&0&0&0&113&0&686&1&179\\
H:2&47&0&0&0&81&0&0&0&151&0\\
I:3&143&135&693&223&1&102&401&0&1&317\\
L:6&99268&0&283&515&461&130&1&593&1&150\\
K:2&0&99246&287&0&0&1&107&3&1&1\\
M:1&53&63&98686&0&0&0&42&2&0&46\\
F:2&363&0&1&98928&2&211&1&2&303&149\\
P:4&104&0&0&1&98949&148&85&0&0&0\\
S:6&96&1&1&221&485&98951&286&91&117&1\\
T:4&0&48&86&0&146&151&98725&0&0&1\\
W:1&18&0&0&0&0&4&0&98760&0&0\\
Y:2&1&1&0&171&0&63&0&2&99034&0\\
V:4&102&1&166&143&1&1&2&2&1&98798\\
\hline
\end{tabular}
\end{center}
\end{table}

\noindent
the diagonal elements of the MPM1 table have
been
defined with the help of the following formula, 

\begin{equation}
M_{jj}=1-\lambda m_j,
\end{equation}

\noindent
and the value of the parameter $\lambda$ has been calculated  from the 
condition that, in the case of 1\% of amino acid substitutions, 
the total fraction of the unchanged amino acids
is equal to

\begin{equation}
\sum_{j=1}^{20} f_j M_{jj}=0.99.
\end{equation}

\setlength{\tabcolsep}{4pt}
\begin{table}
\begin{center}
\caption{Relative mutabilities and fractions of 20 amino acids in the compared sequences. 
We used the convention that mutabilities are relative to alanine and
it is arbitrarily assigned a value of 100.}
\label{mutability}
\begin{tabular}{|l|l|l|}
\hline\noalign{\smallskip}
amino acid & relative mutability ($m_j$) & fraction ($f_j$)\\
\noalign{\smallskip}
\hline
\noalign{\smallskip}
A&	100.00	&0.0449\\
R&	126.09	&0.0369\\
N&	110.42	&0.0671\\
D&	109.29	&0.0579\\
C&	262.91	&0.0073\\
Q&	~77.67	&0.0206\\
E&	~89.88 &0.0644\\
G&	~92.21	&0.0582\\
H&	173.18	&0.0114\\
I&	100.12	&0.0964\\
L&	~75.23	&0.1060\\
K&	~77.51	&0.0930\\
M&	135.01	&0.0184\\
F&	110.12	&0.0690\\
P&	108.02	&0.0238\\
S&	107.78	&0.0796\\
T&	131.05	&0.0333\\
W&	127.44	&0.0047\\
Y&	~99.25	&0.0427\\
V&	123.56	&0.0645\\
\hline
\end{tabular}
\end{center}
\end{table}

\section{Discussion of results}
The major qualitative 
difference between  the MPM matrix introduced in the previous section and the
PAM1 PET91 matrix published in the paper by Jones et al.\cite{l_Jones} is
 that the first one 
is a result of pure mutational pressure whereas the second one is
a result of both mutational and selection pressures. Thus, we have two evolutionary 
mechanisms responsible for the resulting PAM matrices. 

\begin{figure}
\begin{picture}(100,210)(-50,30.0)
\epsfxsize=9.5cm
\epsfysize=11cm
\epsffile{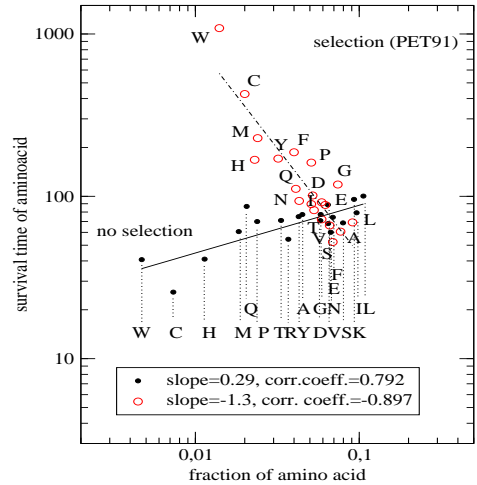}
\end{picture}
\caption{Relation between survival time of amino-acids and their fractions 
in compared pairs of homologous genes in the case with selection  
and in the case without selection. Selection data for amino acids 
have been taken from PAM1 PET91 matrix whereas the data in the case 
without selection 
have been simulated using experimentally found mutational pressure of
the {\it B. burgdorferi} genome. 
}
\label{fig1}
\end{figure}

With the help of formula (\ref{r_approx}) (extended to amino acids) 
we have calculated effective survival times of amino acids in the case 
of the MPM$1$ matrix (Table \ref{table2}) and the mutational/selectional 
PAM$1$ PET$91$ matrix (\cite{l_Jones}). The value of the parameter $\lambda$ is a 
counterpart of  $u$ in (\ref{r_approx}).
In Fig.\ref{fig1}, 
we presented the relation between the calculated survival time 
o amino acids and their fractions in the {\it B. burgdorferi} proteins, in the pairs of diverged genes, in a log-log scale.  
One can observe that the data are highly correlated and in both cases 
the dependence of the mean survival time of amino acid on the fraction of the
amino acid represents a power law:

\begin{equation} 
\tau_j \sim F_j^{\alpha}
\end{equation}

\noindent
with a negative value of $\alpha \approx -1.3$ in the case of selection
and a positive value of $\alpha \approx 0.3$ in the case of mutation
pressure on the leading DNA strand of the {\it B. burgdorferi} genome.
The value of $\alpha$ for the analogous mutational PAM1 
matrix calculated in the case of the lagging DNA strand of the 
{\it B. burgdorferi} genome is about twice as small. It is worth to 
underline that the slopes $\alpha$ are the same 
for the matrices PAM$k$ with high values of $k$,  and thus, they are 
universal with respect to 
evolution.

\begin{figure}
\begin{picture}(100,190)(5,0.0)
\epsfxsize=8.6cm
\epsfysize=10cm
\epsffile{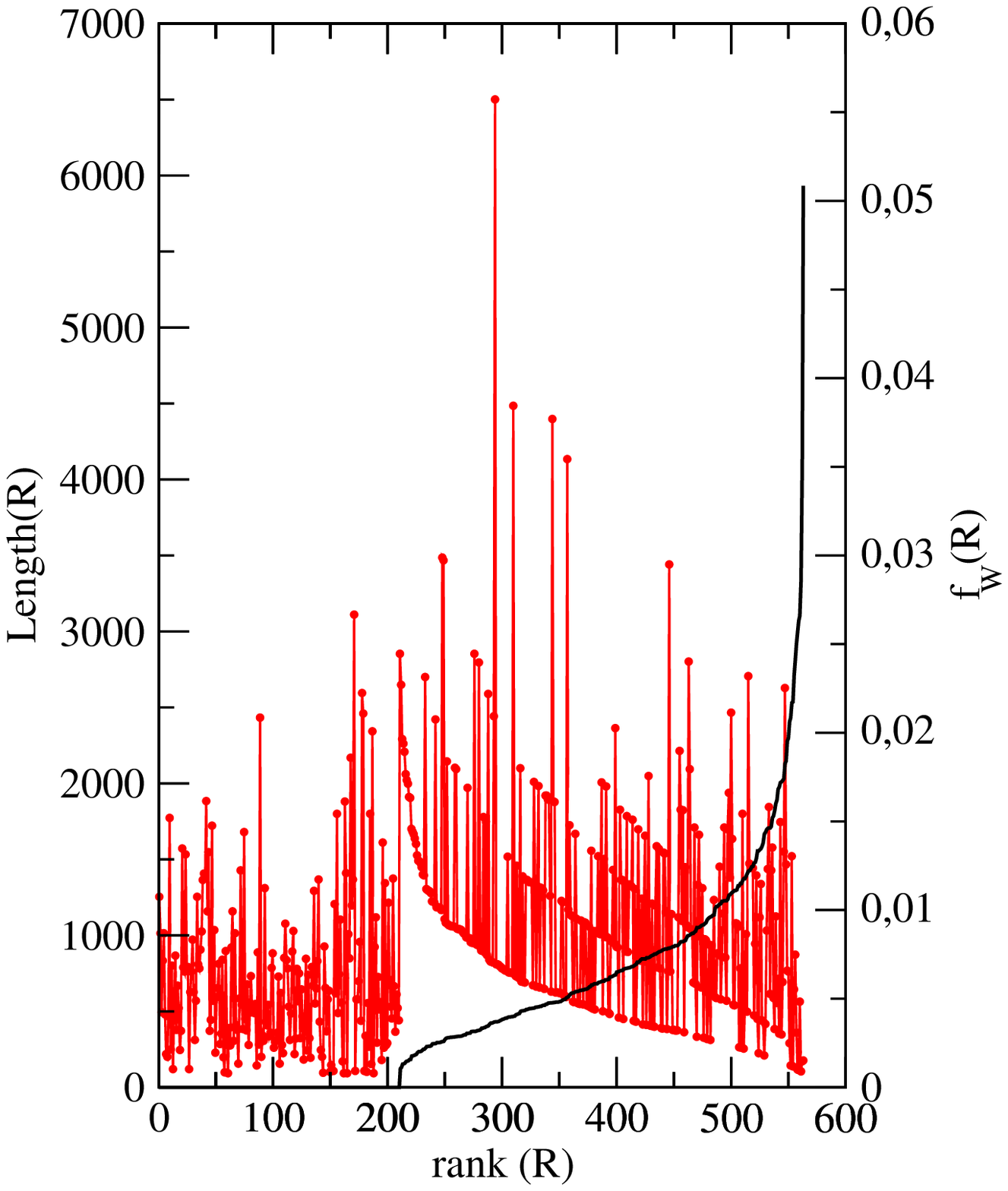}
\hspace{-3cm}
\begin{picture}(100,190)(-10,0.0)
\epsfxsize=7.6cm
\epsfysize=10cm
\epsffile{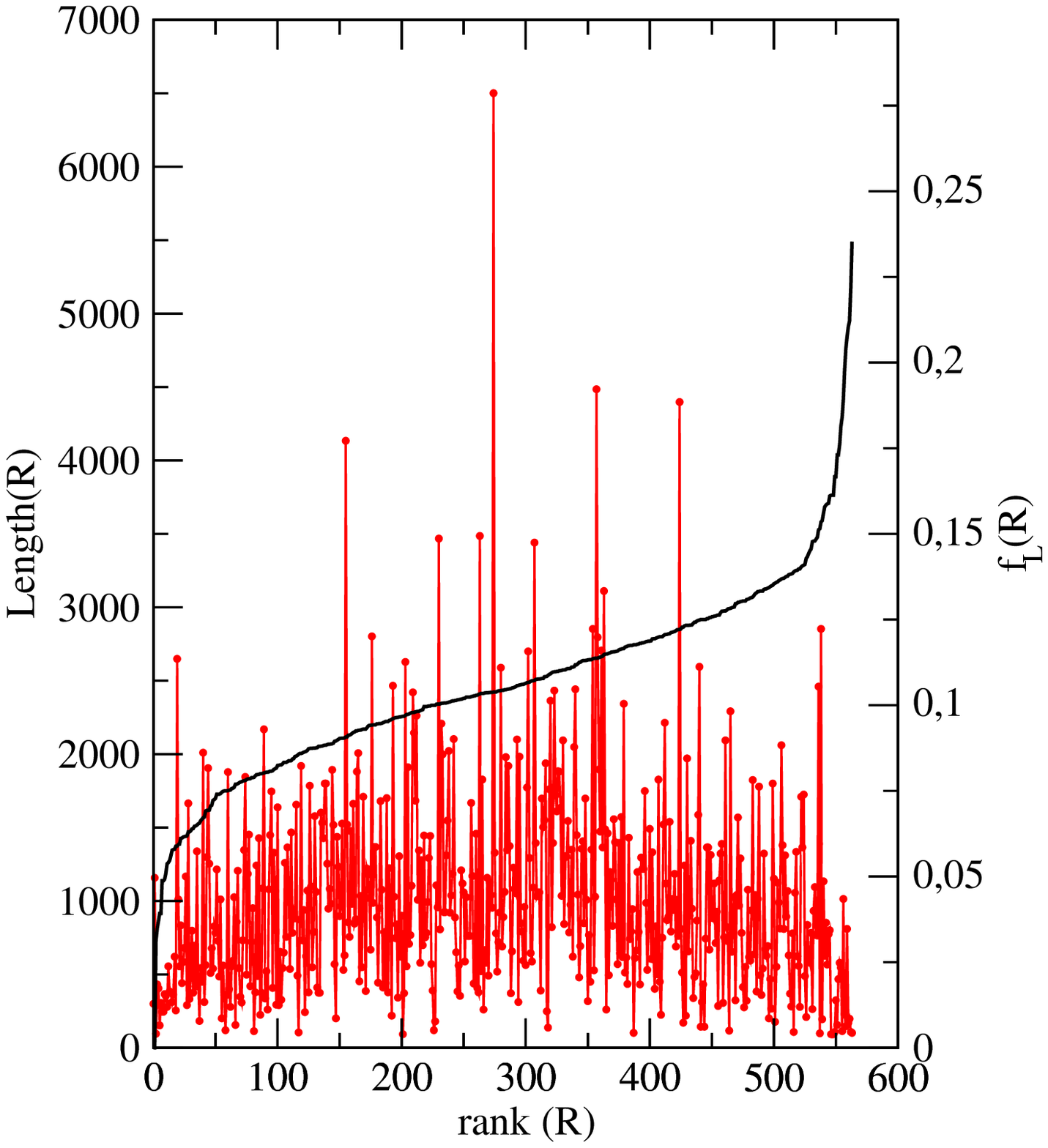}
\end{picture}
\end{picture}
\caption{Gene length versus gene rank, where 
each gene has assigned a rank with respect 
to fraction of tryptophane (left graph) and with respect to leucine 
(right graph). In each graph 
there are two plots: the dots represent gene's size vs. rank, 
whereas the second 
plot represent the fraction vs. rank.}
\label{fig2}
\end{figure}

In Fig.\ref{fig1} we may observe a kind of evolutional scissors
acting on amino acids. Once the less frequent amino acids, like $W$
(tryptophane), $C$ (cysteine),  
have much shorter turn over time compared with 
other amino acids (as can be seen from the lower line)  
the selection pressure (upper line) counteracts with the effect. 
On the other hand, the most mutable amino acids, like $L$ (leucine) or 
$I$ (isoleucine) , which
are very frequent in genes,  
seem to be much weakly influenced by selection.

In genes the fraction of the amino acids most protected 
by selection strongly depends on gene size, it is diminishing 
when gene's size is increasing. The effect weakens  
if we go into right the direction of the evolutionary scissors in Fig.\ref{fig1}.

To show this, we ordered all 564 genes under consideration 
with respect to fraction of an examined amino acid and the genes have been
assigned a rank number. Next, we plotted the dependence of both the gene size on
the rank and the dependence of the amino acid fraction in the gene on the rank.
The resulting plots in Figs.\ref{fig2} correspond to two 
evolutionary extreme cases,
representing  tryptophan and leucine. 
It is evident that in the case 
of tryptophan the fraction of that amino acid in genes is 
anti-correlated with the gene's size 
(notice, that about $1/3$ genes do not posses tryptophan). 
In the case of leucine
there is a crossover and the effect of selection 
is evident only for the genes which have more
than 10\% of that amino acid. When the fraction of leucine is less
than 10\% there is even a reverse effect, i.e., the increasing fraction is
correlated with the increasing gene's size. If we look at the evolutionary
scissors in Fig.\ref{fig1}, we can see that in the case of leucine 
the survival time, which originates from pure mutational pressure 
is longer than its selectional counterpart. Recently, there has 
appeared a paper 
by Xia and Li \cite{l_Xia} discussing which amino acid 
properties (like polarity, isoelectric point, volume etc.) 
affect protein evolution. Thus, there is a possibility to 
relate these properties to our discussion of the evolutionary scissors.

\section{Conclusions}
We have shown that the amino acids which experience the highest selectional
pressure have the shortest turn over time with respect to mutation pressure.
The fraction of these amino acids in genes depends on the gene's size.
Much different is the selectional role of the amino acids, like leucine, from
the right hand side of the selectional scissors. Although they have long 
turn over time with respect to the mutational pressure, their fraction
cannot be too high. This could be considered as an effect of optimisation of the genetic information on coding with processes of mutagenesis and phenotype selection for protein functions.
  
~\\
\noindent{\bf Acknowledgemets}\\ 
The work was supported by the grant number 1016/S/IMi/02.\\

\end{document}